\documentclass[preprint]{revtex4}
\usepackage{epsfig}
\usepackage{graphicx}
\usepackage{psfrag}
\usepackage{setspace}
\usepackage{multirow}
\usepackage{braket}
\begin{document}

\title{A Study of  Pc4-5 Geomagnetic  Pulsations  in the Brazilian Sector.}
\author{D. Oliva$^{1}$, M.C. Meirelles$^{2}$ and A. R. R. Papa$^{1,3}$}
\address{\vskip 1.2cm
$^1$ Observat\'orio Nacional\\
 RJ, 20921-400, Rio de Janeiro, Brazil.\\
$^2$ Schlumberger\\
RJ, 20030-021, Rio de Janeiro, Brazil.\\
$^3$ Universidade do Estado do Rio de Janeiro UERJ,\\
RJ, 20550-013, Rio de Janeiro, Brazil.}

\date{March 21, 2014}

\begin{abstract}
This paper presents a study of {\bf Pc4-5} geomagnetic pulsations illustrated by those which were observed after the sudden commencement of May $02$ of $2010$ at $09:08\,\, UT$ at the Brazilian stations {\bf TTB}, {\bf VSS} and {\bf SMS}. We carry out the spectral analysis of a bivariate data using the Morse wavelets and calculate polarization attributes (ellipticity ratio, tilt angle and phase difference) in the time-frequency domain.  The main pulsation wave packets occurred, for the selected day, around noon and a small enhancement of the pulsation amplitude is observed in the {\bf TTB} station.  A change in the pulsation polarization has been found for the {\bf TTB} station, which we have attributed to effects of the equatorial electrojet.
\end{abstract}
\maketitle

\section{Introduction}

\bigskip Geomagnetic pulsations ({\bf GP}) are the ground signature of ultra-low-frequency hydromagnetic waves ({\bf ULF HW}), which are the result of the complex interaction between the solar wind and the Earth's magnetosphere. The widely accepted approach for the explanation of {\bf GP} is that the magnetosphere acts as resonant cavity and waveguide.  It responds to a broadband stimulus by resonating at specific frequencies, making these cavity modes couple to field line resonances and travel to the ionosphere. Finally, the induced ionospheric currents radiate electromagnetic waves traveling to the Earth surface.

The classification of a {\bf GP} is made based on its morphology and frequency. Firstly they are divided in continuous ({\bf Pc}) and irregular ({\bf Pi}), where the first includes oscillations with quasi-sinusoidal waveforms and the second one pulsations with irregular shape. Regular pulsations are classified, following the oscillation frequency, in five ranges: from {\bf Pc1} to {\bf Pc5} covering the range from $0.2 \,Hz$ to $2 \, mHz$. The irregular ones are grouped into two frequency ranges:  {\bf Pi1} and {\bf Pi2} (from $1\, Hz$ to $2\, mHz$) \cite{1} . Others classifications that extend the frequency range or use as grouping criterion its causality can be found in the literature \cite{2}.

{\bf ULF HW}'s characteristics depend on the values of the main parameters that govern the solar-terrestrial dynamics (interplanetary magnetic field and solar wind pressure), from the magnetosphere regions of origin and from regions through which the trajectory of pulsations to the ground, passes.
Many {\bf ULF HW} are generated during the different phases of substorms, where different kinds of plasma instabilities take place. For instance, during the current wedge formation in the substorm expansion phase, the perturbations, carried out by the field-aligned system, propagate to the ionosphere generating {\bf ULF HW}.  {\bf Pi1-Pi2 HW} have been studied to link the optical and magnetic stages of the substorm phase onset \cite{3}.

GP have been related to the compression from solar wind pressure associated with storm commencements ({\bf SSC}) and sudden impulses ({\bf SI}).  Pc4-5 events, with a sharp power spectra with dominant period of $T\,=\,232\,s$, have been observed during the initial storm phase. During the recovery phase {\bf Pc5 GP} presenting broad band spectra with peaks between $300$ and $500\,s$ have been identified, which have been connected to the starting of partial ring current associated with the {\bf SSC} due to instabilities of hot ring current plasma \cite{4}. It should be noted that there are not {\bf ULF HW} linked specifically to the storm-time ring current. A more plausible explanation links {\bf GP} detected during the storm-time to substorms that take place during magnetic storms.

The {\bf Pc4-Pc5} pulsations can also be originated (or affected) by sudden storm commencements {\bf SSC} and sudden impulses {\bf SI}, although their time lags are difficult to establish because their periods are comparable to the rise time of the {\bf SSC} or {\bf SI}. In general, the pulsation period can be shortened or elongated due to the compression or expansion of the magnetosphere during the {\bf SSC} or negative {\bf SI} events \cite{5}.
{\bf Pc4} and {\bf Pc5} pulsations are characterized by a rather sinusoidal appearance with periods ranging from $45\,s$ to $600\,s$, showing similar behaviors at the  geomagnetic conjugate points.  The amplitude of {\bf Pc4} pulsations is typically lower than for {\bf Pc5} pulsations. Usually the amplitude ratio {\bf Pc4/Pc5} is around one percent. These pulsations are driven by energy transfer from compressional modes, magnetospheric waveguide modes, cavity modes and Kelvin-Helmholtz instabilities at the magnetopause boundary \cite{6}. The latitudinal dependence of {\bf Pc4} and {\bf Pc5} pulsation properties has been largely studied. For higher latitudes it has been found that almost all micropulsation trains show the same frequency at all latitudes. However, both the latitudes of intensity maximum of spectral components and the latitudes where the polarization reversal of the pulsations takes place decrease with increasing frequency \cite{7}.

In this paper we present a comparative investigation of ULF Geomagnetic pulsations ({\bf Pc4-Pc5} range) detected at three Brazilian ground stations Tatuoca ({\bf TTB}), Vasouras ({\bf VSS}) and  S\~ao Martinho da Serra ({\bf SMS}). We illustrated our studies with the storm sudden commencement ({\bf SSC}) occurred on May 02, 2010 at 09:08 UT.
We analysed the two time series formed by the components $Hx$ and $Hy$ of the Earth magnetic field. Employing the Morse wavelets we have calculated some polarization attributes (ellipticity ratio, tilt angle and phase difference) in the time-frequency domain, and using these parameters we have examined the latitudinal dependence of low-latitude {\bf Pc4-5} pulsations in the Brazilian sector.

\section{Data set}
\begin{figure}[[htbp]
\includegraphics[height=6.2cm, width=7.5cm]{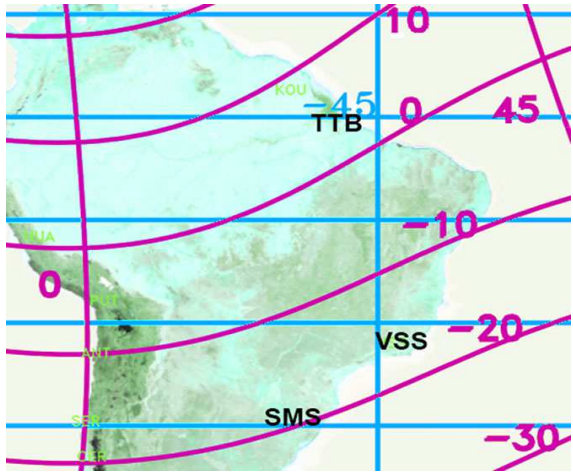}
\caption{\label{EPS_JPGMAX_Geomagnetic_Coordinates_by_SuperMag} Map showing the three Brazilian observatories comprised in this study. Labels indicate the IAGA code of observatories. Their geographic co-ordinates are: S. Martinho da Serra (-29,538, -53,855), Vassouras (-22,404, -43,663) and  Tatuoca (-1,2001, -48,506). The geomagnetic coordinates corresponding to the year 2010 (using IGRF-11) are: S. Martinho da Serra (19.85 S, 16.98 E ), Vassouras (13.43 S, 27.06 E) and  Tatuoca (7.95 N, 23.96 E). Figure edited from $ http://supermag.uib.no/info/img/SuperMAG_Earth_GEO.png)$.}
\end{figure}

The data set consists of a record of $Hx$ and $Hy$ components of the geomagnetic field with a resolution of one second, which was collected during the first five months of  2010 at the Brazilian geomagnetic observatories of Tatuoca ({\bf TTB}), Vassouras ({\bf VSS }) and  S\~ao Martinho da Serra ({\bf SMS}). These observatories comprise almost all the Brazilian zone from mid-latitudes to the dip-equator (see Figure \ref{EPS_JPGMAX_Geomagnetic_Coordinates_by_SuperMag}).

\begin{figure}[htbp]
\includegraphics[height=9.3cm, width=15.5cm]{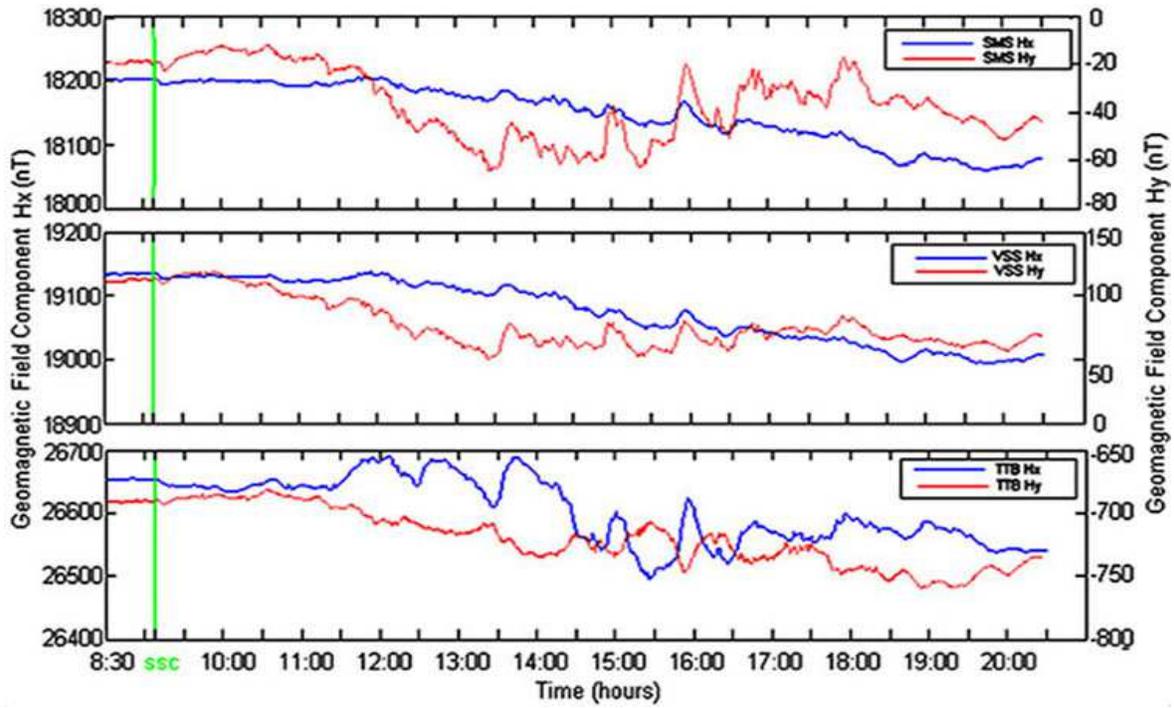}
\caption{\label{Geomag_Field_Comp} Raw magnetic data (Hx, Hy) registered at SMS, VSS and TTB magnetic observatories in May 02, 2010. }
\end{figure}

To illustrated our study, we have taken  the data corresponding to time interval of the day May 02 of 2010 recorded at the three stations to study the pulsations associated to a storm sudden commencement ({\bf SSC}) occurred at 09:08 UT of that day \cite{9}.
Figure \ref{Geomag_Field_Comp} shows the raw geomagnetic records measured on May 02, 2010 at the three stations for the time interval from 8:30 UT to 20:30 UT. The onset of the {\bf SSC} is depicted by a vertical line at 09:08 UT.

\begin{figure}[htbp]
\includegraphics[height=9.3cm, width=15.5cm]{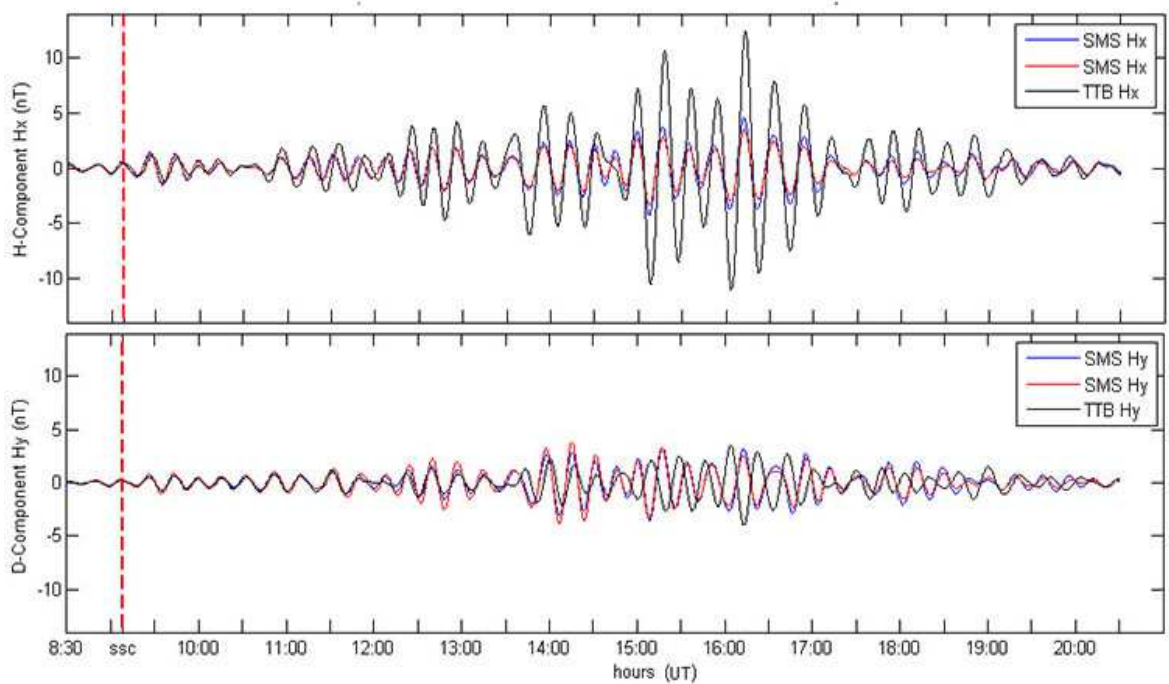}
\caption{\label{Puls_Hx_Hy_para_SMS_VSS_TTB} Filtered data showing the pulsation event recorded at SMS, VSS and TTB magnetic observatories in May 02, 2010. }
\end{figure}

\section{Data processing}
In order to isolate pulsations of interest ({\bf Pc4-5} range), data sets were filtered using a Butterworth bandpass filter (covering the band from 20 to 100 seconds).
Figure \ref{Puls_Hx_Hy_para_SMS_VSS_TTB} displays the filtered data showing the pulsation event of May 02, 2010 recorded at {\bf SMS}, {\bf VSS} and {\bf TTB} stations. The upper and lower  panels show the $Hx$ and $Hy$ components respectively. It can be appreciated around local noon, an enhancement of the pulsation $Hx$ component at {\bf TTB} whereas for the $Hy$ component it is observed a phase shift with respect to the other two stations.

\begin{figure}[htbp]
\includegraphics[height=9.3cm, width=17.5cm]{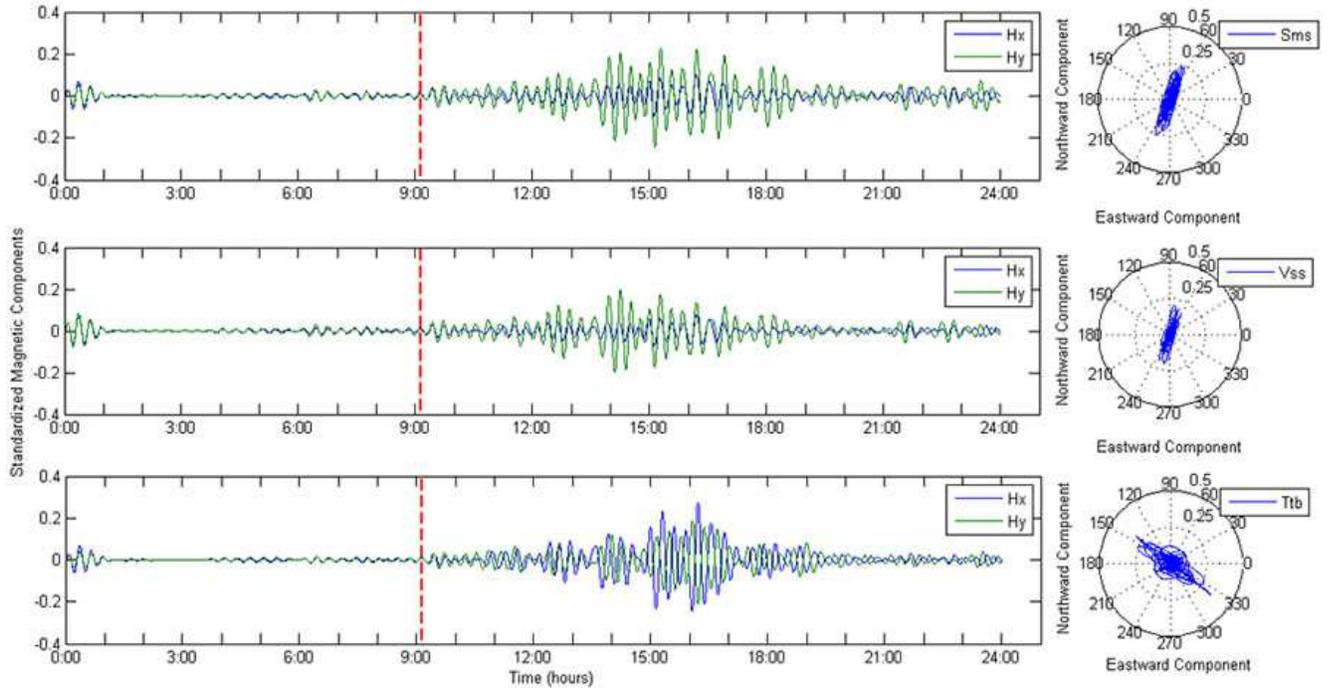}
\caption{\label{Pulsations_and_Hodogramas}  Pulsations at  the three magnetic observatories. From top to botton, panels correspond to $Hx(t)$ and $Hy(t)$ at SMS, VSS and TTB respectively.}
\end{figure}

It has been made the standardization of data sets to obtain a scaled centered version of them. 
Figure \ref{Pulsations_and_Hodogramas} shows the standardized magnetic components of the pulsation and its hodograms.  A well defined pulsation train is noted centered around $15:00 UT$ ($12:00 MLT$) at the three stations.  In Vassouras and S\~ao Martihno da Serra stations it can be observed that the amplitude of the pulsation in the direction {\bf WE} is greater than the {\bf NS} one, but at Tatuoca station this behavior is inverted, while the pulsation amplitude is also enhanced.  Observing the hodograms at the right of Figure \ref{Pulsations_and_Hodogramas} it can be appreciated that the orientation of polarization ellipse changes from north-east for {\bf SMS} and {\bf VSS} stations to north-west for the {\bf TTB} station.
We have shown the hodograms for the standardized data because the above effect is more visible on them. For the non standardized data this effect, although present, is more difficult to appreciate due to the very tight shape of the polarization ellipses.

This results concord with former researches on the ionospheric effects of the {\bf EEJ} on the polarization properties of {\bf ULF} pulsations in low latitudes, which state that the associate ionospheric gradients present in this region act on the amplitude of the D-component on ground. As a consequence, the azimuth of the polarization ellipse exhibits a counterclockwise rotation to northwest, while the ellipticity is subject only to little changes. However it should be pay attention to the fact that ground induction could also alters the polarization characteristics of {\bf Pc4-5} pulsations \cite{Arora}.

\section{Wavelet Spectral Analysis}

Wavelet spectral analysis allows us to quantitatively monitoring  signals evolution by decomposing a time-series into time-frequency space. In this way,  some problems associated to Fourier methods can be overcome, determining both the dominant modes of variability and how those modes vary in time. Wavelets are especially useful for signals that are non-stationary, have short-lived transient components, have features at different scales or have singularities \cite{Torrence}.
In our case we are dealing with two-component signals, thus we are interested in incorporating the time–frequency polarization analysis to determine the oscillation properties of {\bf Pc4-5} pulsations during events.  The polarization analysis can be achieved using progressive wavelets, which are characterized by the fact that the  Fourier coefficients for negative frequencies are zero. As a consequence, the wavelet transform of a  bivariate signal will be equivalent to the bivariated analytical one. This allows us to obtain the ellipse parameters without the use of  bivariate analytical signals, i.e. avoiding the calculation of the Hilbert transform, as occurs in the complex trace method.  In order to implement the above scheme  we have constructed one complex time series ($ C(t)$ ) with the bivariate time series $Hx(t), Hy(t)$ as  $C(t)= Hx(t) + i Hy(t)$. We have applied then the progressive wavelet transform on $C(t)$ and from it we have obtained polarization parameters in the time-frequency domain. In our calculations we have followed the method used in \cite{12}, but in place of the Cauchy Wavelet we have used the Morse wavelet. It is a biparametric wavelet obtained from an autovalue problem, which includes as a particular case  (for the parameters $\gamma = 1,\, K = 0 $) the Cauchy one \cite{13}. One noted advantage of the use of this wavelet is the improvement of statistical estimates working like tapers \cite{14}.

\begin{figure}[[htbp]
\includegraphics[height=9.3cm, width=15.5cm]{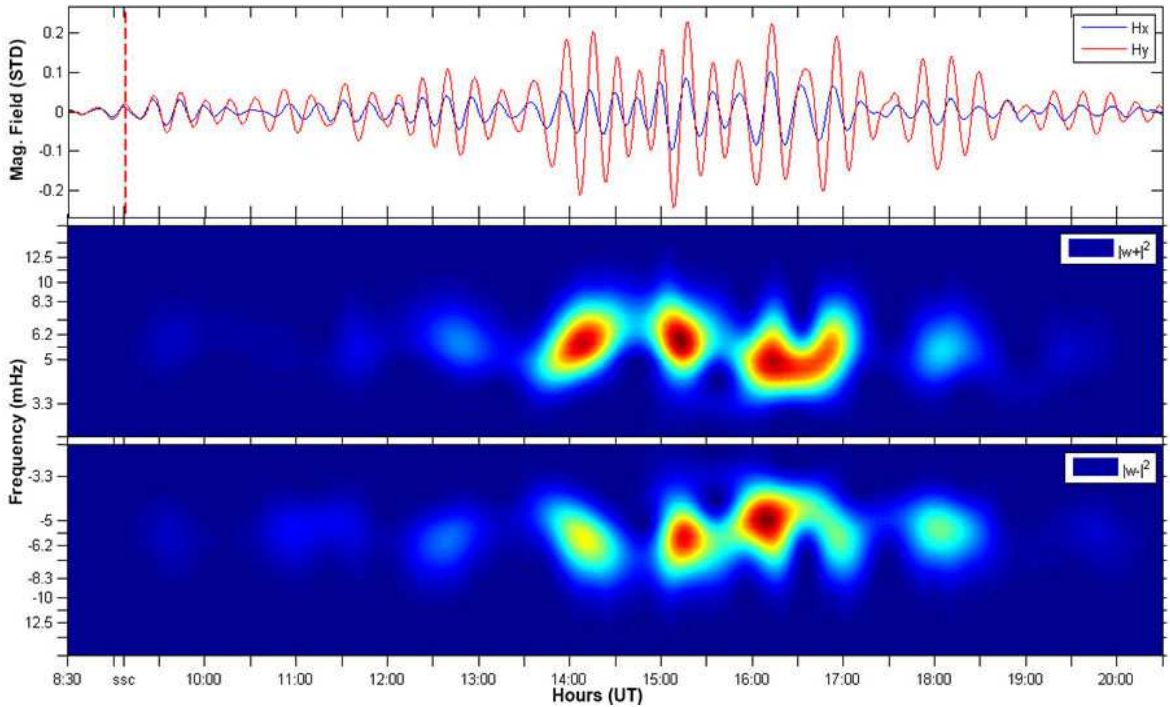}
\caption{\label{Rotatory_Wavelet_Transforms_SMS} Standardized magnetic field (top), Progressive (middle)  and regressive (botton) power spectra of the pulsations at S. Martinho da Serra, for the selected date.}
\end{figure}

\begin{figure}[[htbp]
\includegraphics[height=9.3cm, width=15.5cm]{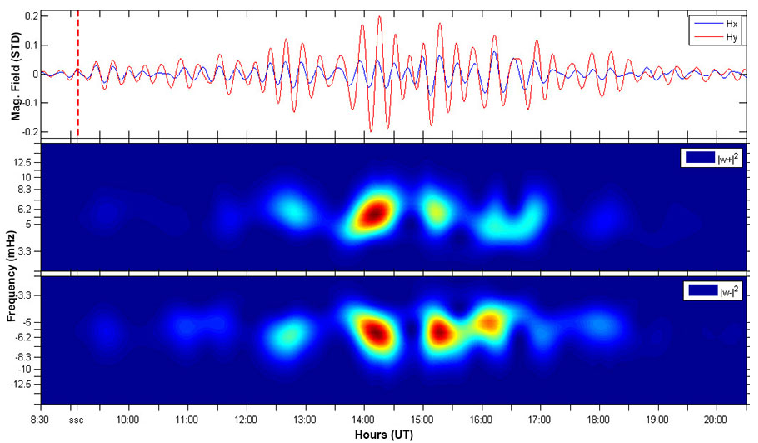}
\caption{\label{Rotatory_Wavelet_Transforms_VSS} Standardized magnetic field (top), Progressive (middle) and regressive (botton) power spectra of the pulsations at Vassouras, for the selected date.}
\end{figure}

\begin{figure}[[htbp]
\includegraphics[height=9.3cm, width=15.5cm]{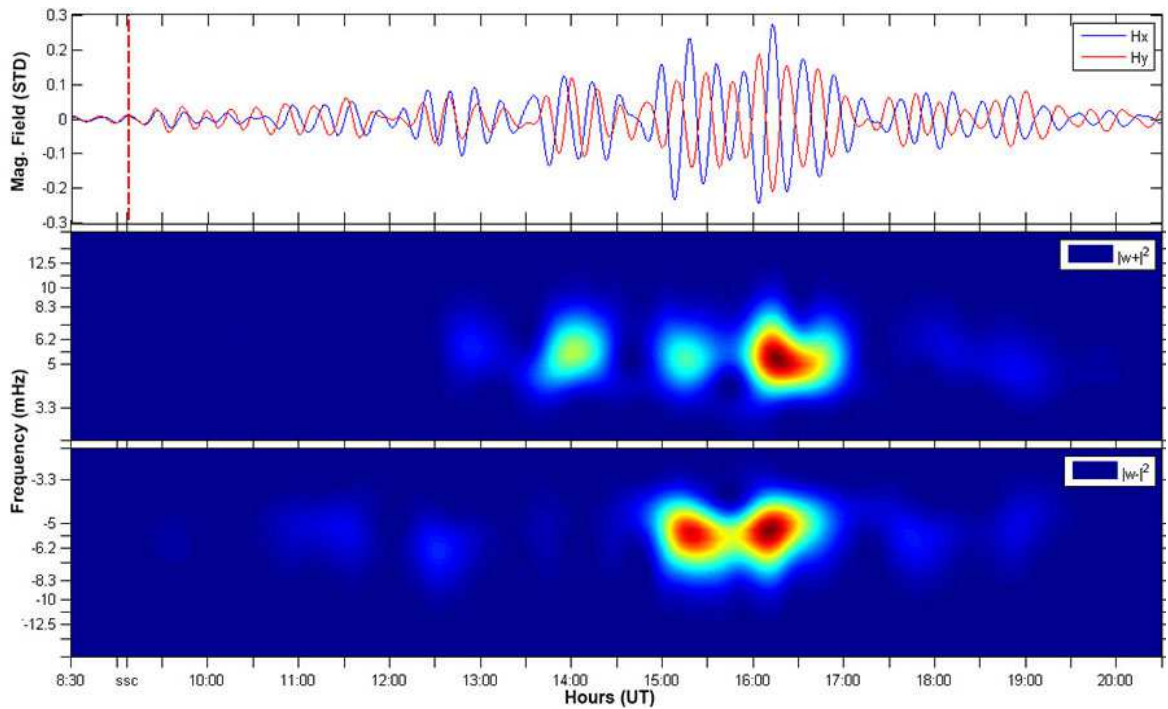}
\caption{\label{Rotatory_Wavelet_Transforms_TTB} Standardized magnetic field (top), Progressive (middle)  and regressive (botton) power spectra of the pulsations at Tatuoca, for the selected date.}
\end{figure}

We have adapted the code JLAB, used in oceanographic research (see \cite{15}), to apply it for the wavelet spectral analysis of geomagnetic pulsations.
The wavelet transform ({\bf WT}) of our data was carried out with the Morse parameters ($ \gamma=3,\, \beta=3 $) and it was generated a logarithmic Morse space to cover the frequency range  from $0.025$ Hz  to $0.002$ Hz ($40$ s to $500$ s). Figures \ref{Rotatory_Wavelet_Transforms_SMS}, \ref{Rotatory_Wavelet_Transforms_VSS} and \ref{Rotatory_Wavelet_Transforms_TTB} show progressive and regressive power spectra of the pulsation trains at the three stations for the selected date.  For each station  the filtered and standardized geomagnetic components $Hx$ and $Hy$ are plotted, where the {\bf SSC} is indicated with a dashed red line. In these figures it can be appreciated that the major contribution to the spectral power is located in the frequency band from $4.1\,mHz$ to $7.3\,mHz$ ($136\,s$- to $200\,s$), which belongs to the frequency zone containing the lower and upper bounds of {\bf Pc4} and {\bf Pc5} pulsations respectively. The pulsations trains are enclosed in the time interval from the 14:00 to 17:00 hour ({\bf UT}).

\section{Polarization Analysis}

Significant information on the generation and propagation mechanisms of {\bf ULF} waves can be obtained from their polarization characteristics.
In the magnetosphere two principal modes are expected to propagate, which can be decoupled from the wave equation for the cold magnetoplasma. Putting the background magnetic field in the $ \hat{z}$ direction, the first mode (the toroidal or guided Alfv\'en mode) can be obtained assuming that the wave number $m\,=\,0$. This mode is polarized in the $\hat{x_2}$ direction and produces standing waves along the field lines due to the ionosphere boundary conditions. The other mode (Poloidal mode) can be isolated assuming that the azimuthal wave number (azimuthal gradient) is large. This mode is polarized in the $ \hat{x_1}$ direction (field aligned direction). In general these modes are coupled in a nonuniform plasma and their polarizations are functions of the background magnetic field \cite{Southwood}.

The concept of polarization states were introduced initially for plane waves in optics. Afterward, the concepts of polarization and degree of polarization were generalized to n-variate time series using the expansion of the spectral matrix in {\em pure states} by means of the calculation of their eigenvalues \cite{Samson1}.

\begin{figure}[[htbp]
\includegraphics[height=4.9cm, width=12.75cm]{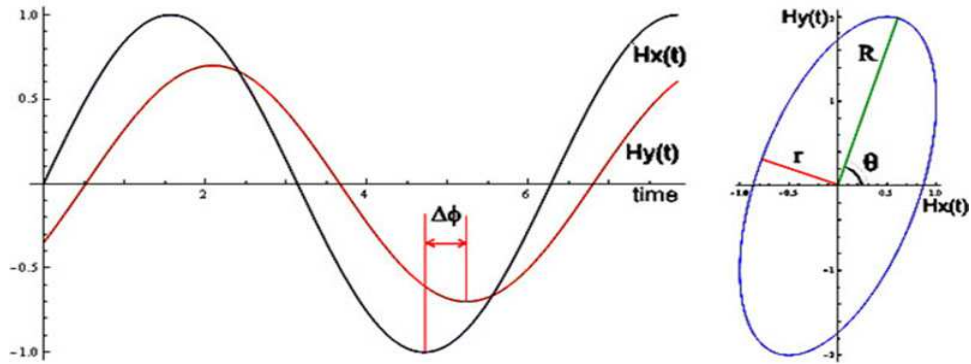}
\caption{\label{Ellipse_Parameters} Illustration of the polarization ellipse and ellipse parameters.}
\end{figure}

The polarization can be described using the polarization ellipse  and its parameters \cite{Born and Wolf}. In figure \ref{Ellipse_Parameters}  the polarization parameters are shown: $ R $ (the semi-major axis $R \geq  0$), {\bf r}  the semi-minor axis $ r \geq 0 $),   $\rho = r/R $ (the ellipticity ratio), $ \rho \in [0; 1] $,   $ \Theta $ (the tilt angle, which is the angle of the semi-major axis with the horizontal axis, $ \Theta \in [-1/2, 1/2] $) and $ \Delta \phi $ (the phase difference between the {\bf x}  and {\bf y} components).

 We have oriented the coordinate system along the Hx magnetic component ( $C(t)= Hx(t) + i Hy(t) $), which corresponds to a left-handed coordinate system. In this case positive and negative phase difference indicates left-handed (l.h) polarization and right-handed (r.h) respectively.

We have calculated, using the progressive and regressive wavelet transforms, the polarization attributes in the time-frequency domain. First we have used a synthetic signal in order to check the changes we have introduced in the above mentioned method.

The values with lower spectral power do not contain useful information about the signal polarization and make difficult  the understanding of plots. Thus, in what follows we display only the values of the elliptic parameters corresponding to a spectral power above given value. We have chosen as cut-off  value the module of the maximum value of the spectral power multiplied by a threshold parameter $\epsilon \, = 0.4 $.

\begin{figure}[[htbp]
\includegraphics[height=9.3cm, width=15.5cm]{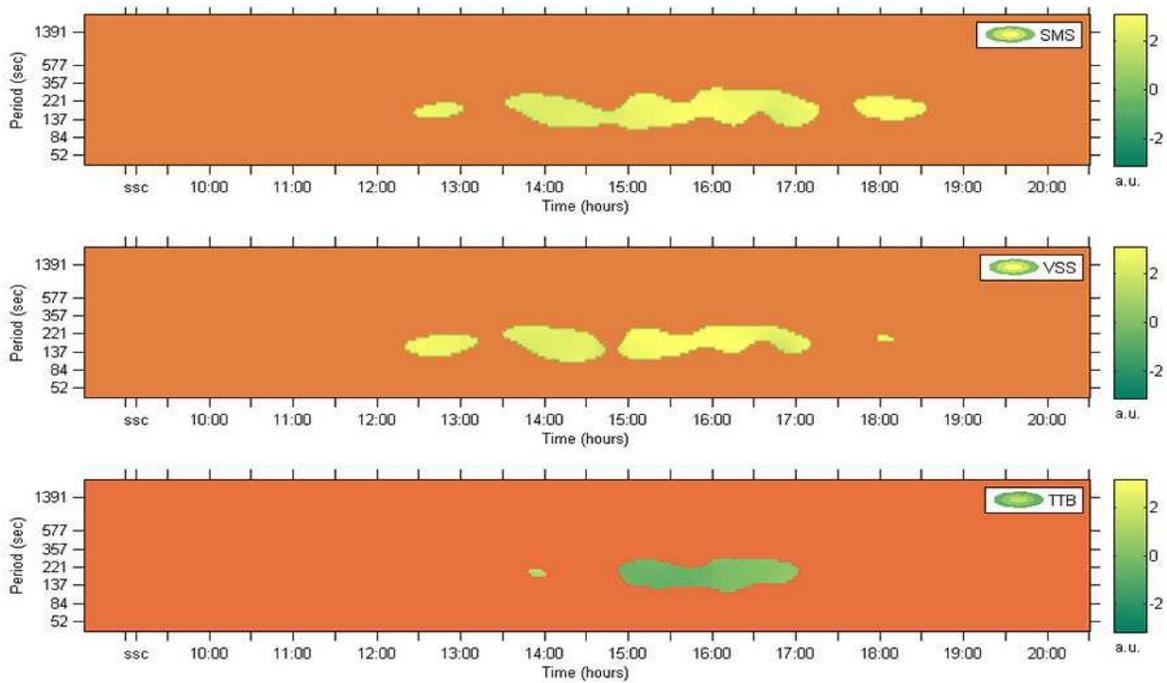}
\caption{\label{Phase_Difference} Phase difference in the time-frequency domain for the pulsation at the three stations, in the selected date, with threshold parameter $\epsilon = 0.4$.}
\end{figure}

Figure \ref{Phase_Difference} shows the phase difference in the time-frequency domain. Here it can be seen that the zone of significative power content is located around the local noon and that the pulsations for the stations of Vassouras and S. Martinho da Serra show a positive phase difference at both sides of the local noon, indicating a left-handed ({\bf l.h.}) elliptical polarization. However for the Tatuoca station it can be observed a small region of positive phase difference in the prenoon sector, while in the postnoon sector the phase difference becomes negative and almost zero showing a right-handed ({\bf r.h.}) quasi-linear polarization.

\begin{figure}[htbp]
\includegraphics[height=9.3cm, width=15.5cm]{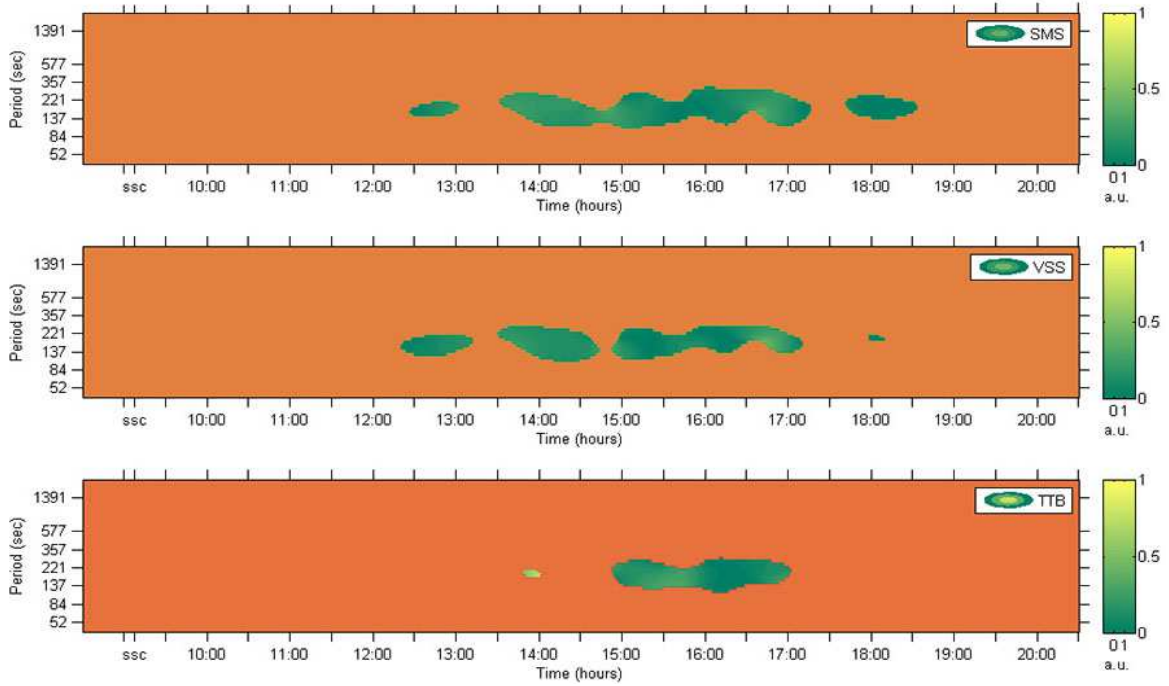}
\caption{\label{Ellipticity} Ellipticity ratio in the time-frequency domain for the three stations, in the selected date, with threshold parameter $\epsilon = 0.4$.}
\end{figure}

The ellipticity ratio computed in the time-frequency domain is displayed in Figure \ref{Ellipticity}. It shows non zero small values for all stations, which says that the polarization ellipse has a tight shape, i.e., the oscillations are concentrated in the direction defined by the tilt angle. It can be also appreciated that the ellipticity does not experiment an appreciable daily variation. It is in accordance with former results on Pc4 pulsations, which assert that the ellipticity does not suffer  appreciable changes by the sunrise effect \cite{Saka}.
The fact of ellipticity at {\bf TTB} station not displaying noticeable changes with respect to {\bf SMS} and {\bf VSS} stations says that this parameter is not sensitive to the influence of the {\bf EEJ}.

\begin{figure}[htbp]
\includegraphics[height=9.3cm, width=15.5cm]{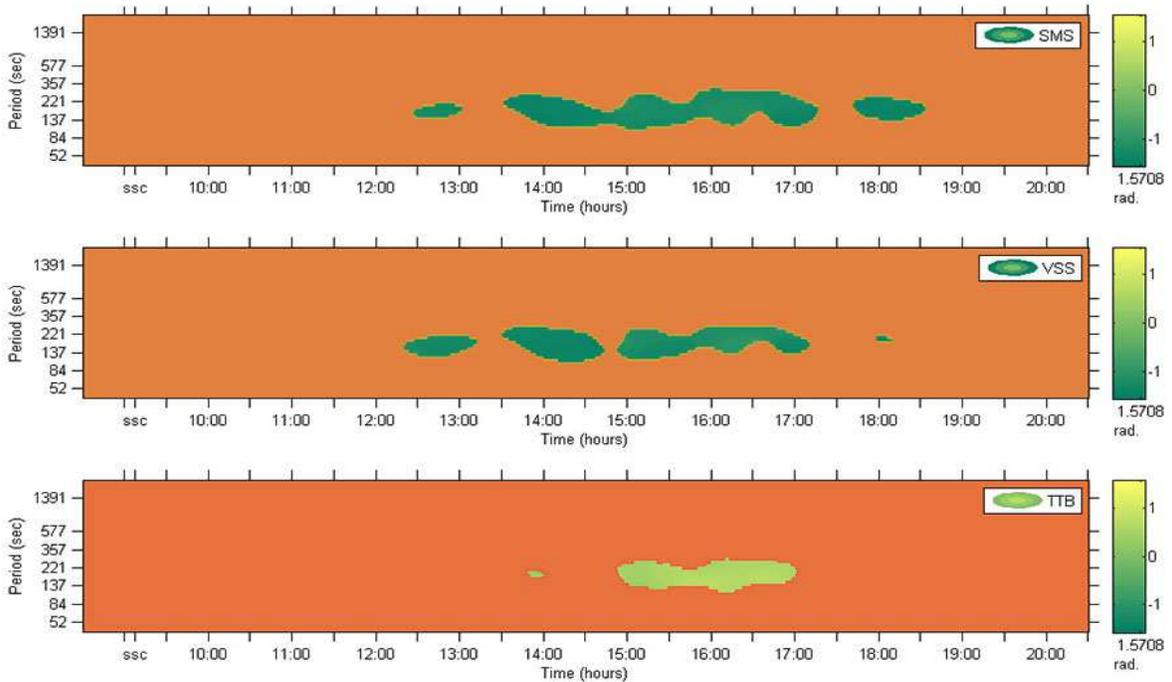}
\caption{\label{Tilt_Angle} Tilt angle  in the time-frequency domain for the three stations, in the selected date, with threshold parameter $\epsilon = 0.4$.}
\end{figure}

Figure \ref{Tilt_Angle} displays the tilt angle in the time-frequency domain. It shows negative values for Vassouras and S. Martinho da Serra and positive for Tatuoca. These values of tilt angle validate the left-handed character of the pulsation for the first two stations, whereas the right-handed ({\bf r.h.}) or linear for the last one. This result is in concordance with the "rotation effect" observed for {\bf Pc3-4} pulsations, which consists in the rotation of the major axis of the polarization ellipse without affecting the ellipticity \cite{Saka}. In our case this effect is observed only at {\bf TTB} station, thus, we associate it to conductivity ionospheric gradients produced by the {\bf EEJ}.

\begin{figure}[htbp]
\includegraphics[height=8.0cm, width=16.5cm]{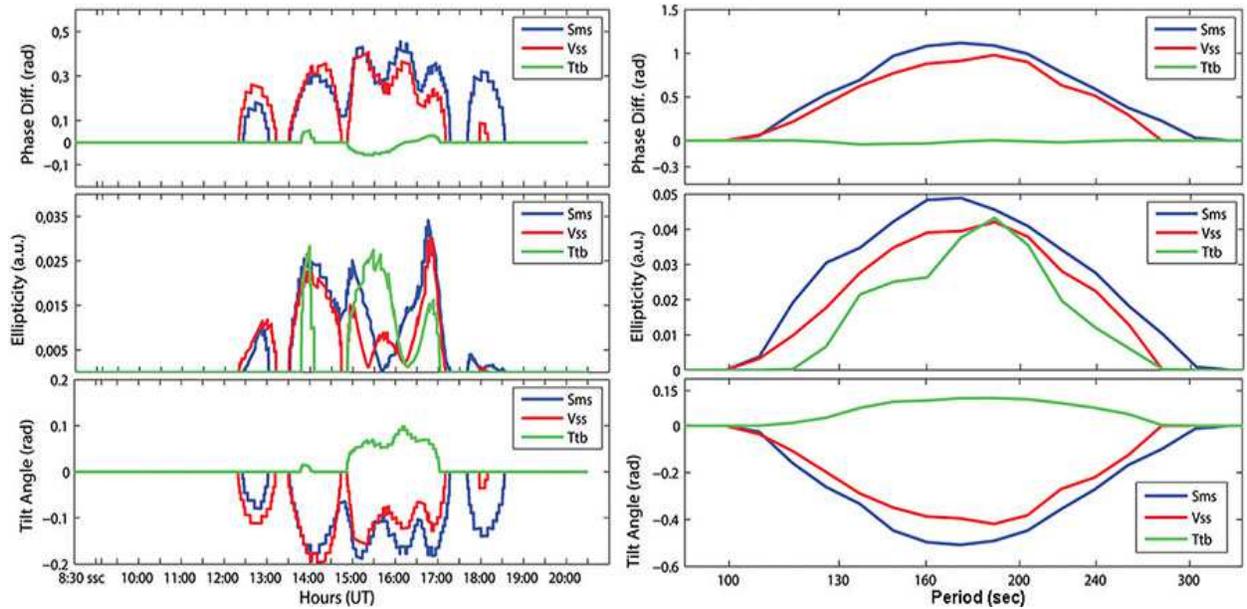}
\caption{\label{AverageEllipseParam} Ellipse parameters averaged along the time and frequency for the three stations, in the selected date.}
\end{figure}

The study of the elliptic parameters in the time-frequency domain becomes more clear using average quantities. Left and right panels of Figure \ref{AverageEllipseParam} show the ellipse parameters averaged in frequency and in time respectively. The averages have been calculated for intervals of significative power spectrum. The left-upper panel of Figure \ref{AverageEllipseParam} displays the phase difference averaged in frequency ($\braket{\Delta \phi}_f$) at the three observatories. It can be observed that $\braket{\Delta \phi}_f$ at {\bf SMS} is slightly smaller than at {\bf VSS} in the prenoon sector, but around noon occurs an inversion, such that in the postnoon sector $\braket{\Delta \phi}_f$ at {\bf SMS} becomes larger than $\braket{\Delta \phi}_f$ at {\bf VSS}. At {\bf TTB} $\braket{\Delta \phi}_f$ exhibits a small positive peak at the prenoon sector, around noon takes negative values and at the postnoon sector returns again to small positive values. The left-middle panel of Figure \ref{AverageEllipseParam} shows the ellipticity averaged in frequency ($\braket{\rho}_f$). It can be observe that this parameter exhibits a similar response at the three observatories and doesn't display any change of behavior around noon. The left-lower panel of Figure \ref{AverageEllipseParam} displays the tilt angle averaged in frequency ($\braket{\theta}_f$). Here it can be observed that $\braket{\theta}_f$  is negative at {\bf SMS} and {\bf VSS}, whereas at {\bf TTB} takes positive values. It can be noted that $\braket{\theta}_f$ at {\bf SMS} and {\bf VSS} exhibits a similar  behavior around noon as occurred for $\braket{\Delta \phi}_f$. In the right-upper panel the phase difference averaged in time ($\braket{\Delta \phi}_t$) exhibits at {\bf SMS} and {\bf VSS} similar positive values, remaining {\bf SMS} slightly larger than at {\bf VSS}  for all the significant frequencies. However $\braket{\Delta \phi}_t$ shows small negative values at {\bf TTB}. In the right-middle panel $\braket{\rho}_t$ shows a similar behavior at the three observatories fulfilling the  relationship $\braket{\rho_{SMS}}_t \,>\, \braket{\rho_{VSS}}_t \,>\, \braket{\rho_{TTB}}_t $ for all significative frequencies. In the right-lower panel it can be observed, that $\braket{\theta}_t$ takes negative values at {\bf SMS} and {\bf VSS}, while at {\bf TTB} $\braket{\theta}_t$ takes positive values for all the significant frequencies. Also, it can be seen that the absolute values of the angles holds  $|\braket{\theta_{SMS}}_t| \,>\, |\braket{\theta_{VSS}}_t|\,>\, |\braket{\theta_{TTB}}_t| $.

The diurnal pattern in the pulsation polarization and phase structure, like shown in Figure \ref{AverageEllipseParam}, has been observed for low-latitude {\bf Pc5} pulsations in the Australian sector  \cite{Ziesolleck93}. Using a two dimensional array of stations the authors detected an increasing phase with increasing longitude in the local morning, small longitudinal phase variation around local noon, and decreasing phase with increasing longitude in the local afternoon. This behavior matches with the pattern shown in the left-upper panel of Figure \ref{AverageEllipseParam} by {\bf SMS} and {\bf VSS}. However only {\bf TTB} exhibits polarization reversal at local noon. In the above mentioned investigation the authors related this diurnal variation of polarization attributes to ionospheric effects. The fact, that the polarization parameters $\Delta \phi$ and $\theta$ obey specific pattern respect to local noon except the ellipticity, has been also reported in an earlier research for {\bf Pc3-4} in the Japan sector \cite{Saka}. In this investigation the authors have attributed such behavior to the ionospheric electron density enhancement due to the sunrise.

\section{Summary and Conclusions}

We have studied the behavior of the geomagnetic {\bf Pc4-5} pulsation recorded in the Brazilian geomagnetic observatories of Tatuoca ({\bf TTB}), Vassouras ({\bf VSS}) and S\~ao Martinho da Serra ({\bf SMS}) associated to {\bf SSC} events, illustrating our results with the {\bf SSC} of May $02$, $2010$. We have detected pulsation wave packets around noon at the three stations.  The pulsations at {\bf VSS} and {\bf SMS} show similar spectral characteristics, whereas the pulsation at {\bf TTB} station presents an amplitude enhancement.  This equatorial enhancement of the ({\bf ULF}) pulsations has been reported in former investigations, e.g., in the American sector were investigated the amplitudes and polarization characteristics of Pc5 pulsations connected with the {\bf SSC} followed by the severe magnetic storm of March 24, 1991. The authors studied the latitudinal response of these pulsations and found, apart from the peak near the auroral oval, another peak in the pulsation amplitude at the dip equator \cite{Trivedi}. Also, in a study of {\bf Pc4}  pulsations in the African sector it was found an intensification of the amplitude of the H component at the dip equator \cite{Takla}.

In the calculation of polarization attributes using methods based on the spectral matrix there is always involved a smoothing process, thus the calculated polarization parameters are average quantities over a given frequency window. The method used here allowed us to calculate the  instantaneous polarization attributes in the time-frequency domain.

The fact that we have detected only at {\bf TTB} station this reversion of the polarization sense at local noon, could be attributed to the ionospheric enhancement of the E layer due to the EEJ.

Local time polarization reversals has been also detected for Pc4-5 pulsation at high latitudes and have been associated with Kelvin Helmholtz Instabilities generated at the magnetosphere boundary \cite{Samson72} and Pc5 at low latitudes and to compressional modes or cavity resonances trapped in the magnetosphere \cite{Ziesolleck93}. In general, all excitation mechanisms driven by the solar wind at the day-side magnetosphere show similar diurnal behaviors of polarization attributes.
The fact of having detected this polarization reversal only at {\bf TTB} could be because the ionospheric conductivity gradients generated by the {\bf EEJ} are stronger than those produced by the normal day-side ionospheric enhancement caused by the increment of the solar radiation.

We are aware that it is difficult to carry out a study on the latitudinal dependence of pulsations using only a small numbers of sparse permanent stations. Usually this kind of investigations have been achieved within campaigns where temporary array of magnetometers are installed in order to cover with more resolution large distances. In spite of this fact we have shown that some information can be extracted processing the data from these permanent geomagnetic observatories. The inclusion of more geomagnetic observatories is advised for future studies.

\begin{acknowledgments}

We acknowledge SuperMAG initiative for the maps available at the website $http://supermag.uib.no/info/img/SuperMAG_Earth_GEO.png$.
This material is based upon work supported by the {\bf DTI-PCI} fellowship $N.380.739/09-7$ of Brazilian Science Funding Agency. A. R. R. Papa. wishes to acknowledge {\bf CNPq} (Brazilian Science Foundation) for a productivity fellowship and {\bf FAPERJ} (Rio de Janeiro State Science Foundation) for partial support.

\end{acknowledgments}

\end{document}